\documentclass[12pt]{iopart}
\expandafter\let\csname equation*\endcsname\relax 
\expandafter\let\csname endequation*\endcsname\relax 

\usepackage{amsmath,amsthm,amssymb}
\usepackage{graphics}
\usepackage{graphicx}
\usepackage{subfigure}
\usepackage{dcolumn}

\newcommand{\be}{\begin{eqnarray}}
\newcommand{\ee}{\end{eqnarray}}
\newcommand{\beq}{\begin{eqnarray}}
\newcommand{\eeq}{\end{eqnarray}}
\newcommand{\pd}{\partial}

\newcommand{\nn}{\nonumber}

\newcommand{\dalm}{\kern1pt\vbox{\hrule height 0.9pt\hbox{\vrule width 0.9pt\hskip 2.5pt\vbox{\vskip 5.5pt}\hskip 3pt\vrule width 0.3pt}\hrule height 0.3pt}\kern1pt}

\begin{document}
\title{Naked singularity explosion in higher-dimensional dust collapse}
\author{${}^{1,2}$Masahiro Shimano and ${}^{3,2}$Umpei Miyamoto}
\address{
${}^{1}$Jumonji Junior and Senior High School, Toshima, Tokyo 170-0004, Japan
\\${}^{2}$Department of Physics, Rikkyo University, Toshima, Tokyo 171-8501, Japan
\\${}^{3}$Research and Education Center for Comprehensive Science, Akita Prefectural University, Akita 015-0055, Japan}

\ead{${}^{1,2}$mshimano@rikkyo.ac.jp}
\ead{${}^{3,2}$umpei@akita-pu.ac.jp}
\date{\today}
\begin{abstract}

In the context of the large extra dimensions or TeV-scale gravity, it has been argued that 
an effective naked singularity, called the visible border of spacetime, 
would be generated by high-energy particle collisions. 
Motivated by this interesting possibility, we investigate a particle creation by 
a naked singularity in general dimensions, adopting a spherically symmetric self-similar dust collapse as 
the simple model of a naked singularity formation. 
The power and energy of the particle emission behave in two distinct ways depending on a parameter in the model. 
In a generic case, the emission power is proportional to the quadratic inverse of the remaining time to the Cauchy horizon, 
which has been known for the four-dimensional case in the literature. 
On the other hand, in a degenerate case the emission power is proportional to the quartic inverse of the remaining time to the Cauchy horizon, 
and depends on the total mass of a dust fluid in spite that the central region of the collapse is scale-free due to the self-similarity. 
In the both cases, within a test-field approximation the energy radiated before any quantum gravitational effect dominates amounts to TeV. 
This suggests that a backreaction is not ignorable in the TeV-scale gravity context, in contrast to the similar phenomena in stellar collapse.

\end{abstract}

%

\maketitle


\section{Introduction}

Can we detect the sign of extra dimensions in a collider experiment?
In the scenarios of large or warped extra dimensions~\cite{ArkaniHamed:1998rs,Randall:1999ee}, 
a possibility has been argued that effective naked singularities, called the visible borders of spacetime~\cite{Nakao_2010,Okawa:2011fv}, and mini-black holes~\cite{Argyres:1998qn} are produced in high-energy particle collisions.

The argument in \cite{Nakao_2010} is simple. 
In the collider experiment, colliding particles can have an impact parameter $b$ which is smaller than the Schwarzschild radius $r_{\mathrm{eh}}$ determined by the center-of-mass energy $M$ of the collision. If this situation is realized, a mini-black hole can be produced. 
On the other hand, if the colliding particles have $b$ larger than $r_{\mathrm{eh}}$ but $M$ is order of the Planck energy $M_P$, the horizon will not form in spite that the curvature radius becomes the order of Planck length. 
In the latter case, the trans-Planckian domain of spacetime can be visible to outer observers, that is a realization of the visible borders of spacetime.

Here, we should mention that Okawa et al.~\cite{Okawa:2011fv} showed by a fully general relativistic simulation that the trans-Planckian domains of spacetime not covered by horizons are produced in the course of black-hole collision in 5 dimensions, which strongly supports the argument of Ref.~\cite{Nakao_2010}.

It is expected that black holes created will thermally decay via the Hawking radiation~\cite{Hawking:1974sw}, which can be detected in the LHC~\cite{Kanti:2008eq}. 
If a naked singularity also radiates quanta like black holes, it can be detected in the collider experiments. 
Semiclassical effects during a naked singularity 
formation have been studied in four dimensions~\cite{Ford:1978ip, Hiscock:1982pa, Brave_1998, Vaz:1998gd, Singh:2000sp, Harada:2000me, Miyamoto:2003wr, Miyamoto:2004ba, Harada:2001nj}. 
Typically, the power of the particle emission diverges at the Cauchy horizon unless the singularity is too week. 
Thus, one can expect similar phenomena in higher dimensions, which have been studied by the present authors and their collaborator in the null dust case~\cite{Miyamoto_Nemoto_2011a, Miyamoto:2010vn}.

The evaluation of the particle creation in the visible border or naked singularity formation is accompanied by some difficulties. Firstly, the process of the naked singularity formation in particle collisions will be a highly asymmetric phenomenon. 
Secondly, when the singularity is globally naked, we have to impose the boundary 
conditions on quantum fields at the singularity, which cannot be done {\it a priori}. 

In the previous work~\cite{Miyamoto:2010vn}, as a first step, we considered the Vaidya spacetime, which describes the spherical collapse of a null dust fluid (i.e., lightlike pressureless fluid), 
and the formation of a {\it locally} naked singularity that makes it possible to estimate the spectrum by avoiding the necessity to impose boundary conditions at the singularity. 
This model, however, needed a fine-tuning to realize the background geometry. 
Therefore, we considered the more generic situation in the last paper~\cite{Miyamoto_Nemoto_2011a}, still adopting the Vaidya spacetime. 
In this paper, as the next step, we generalize the analysis in the last paper~\cite{Miyamoto_Nemoto_2011a} to the Lema\^itre-Tolman-Bondi (LTB) spacetime, which describes the spherical collapse of an inhomogeneous dust fluid (i.e., pressureless timelike fluid). While the LTB model is still highly idealized one, this model provides us a non-trivial generalization that still allows completely analytic calculations. We believe that the analysis in this paper would be a footstep to the analyses of more realistic or complicated models.

This paper is organized as follow. In Sec.~\ref{sec:higher-LTB}, we introduce the higher-dimensional Lema\^itre-Tolman-Bondi solution. 
In Sec.~\ref{sec:map_of_null}, we derive the map of null rays, which plays a key roll for the estimate of the emission power using the geometric-optics approximation. 
In Sec.~\ref{sec:power_and_energy}, we evaluate the power and energy of the particle creation. We devote Sec.~\ref{sec:conclusion} to the conclusion and discussions. Several calculations are relegated to Appendices. We use the Planck units $c=\hbar=G=1$, unless denoted explicitly.

\section{Dust collapse in higher dimensions}\label{sec:higher-LTB}

\subsection{Higher-dimensional Lema\^itre-Tolman-Bondi solution}

The line element of $d = (n+3)$-dimensional ($n \geq 1$) marginally bound Lema\^itre-Tolman-Bondi (LTB) solution in comoving coordinates~\cite{Ghosh_2001} is given by
	\begin{equation}\label{eq:high-LTB}
	 	 ds^2=-dt^2+ \left(\partial_r \tilde{R}(t,r)\right)^2 dr^2+\tilde{R}^2(t,r)d\Omega_{n+1}^2,
	\end{equation}
where $d\Omega_{n+1}^2$ is the line element of a unit $(n+1)$-sphere.
The energy-momentum tensor of a dust fluid is given by 
	\begin{equation}
	 T^{\mu\nu}=\epsilon u^\mu u^\nu,
\;\;\;\;\;
	 \epsilon=\frac{(n+1)\partial_r F(r)}{8\pi \tilde{R}^{n+1}\partial_r \tilde{R}},
	\end{equation}
where $u^\mu$ is the four-velocity of the fluid. $F(r)$ is an arbitrary function, referred to as the mass function. From the Einstein equation, it is found that the circumferential radius $\tilde{R}$ satisfies
	\begin{equation}\label{eq:dotR}
	 \left(\partial_t \tilde{R} \right)^2=\frac{F(r)}{\tilde{R}^n}.
	\end{equation}
Hereafter, we assume $\partial_t \tilde{R}<0$ to obtain a collapse solution 
and the energy density $\epsilon$ to be non-negative everywhere. 
Eq.~(\ref{eq:dotR}) is integrated to give
	\begin{equation}\label{eq:t}
	 t-t_s(r)=-\frac{2}{n+2}\frac{\tilde{R}^{(n+2)/2}}{\sqrt{F}},
	\end{equation}
where $t_s(r)$ is a constant of integration.
Without loss of generality, we can rescale the radial coordinate $r$ to satisfy
	\begin{equation}
	 \tilde{R}(0,r)=r.
	\end{equation}
Then, $t_s(r)$ is given by
	\begin{equation}\label{eq:tc}
	 t_s(r)=\frac{2}{n+2}\frac{r^{(n+2)/2}}{\sqrt{F}}.
	\end{equation}
	
Now, let us assume the spacetime to be self-similar (or homothetic) for simplicity
\footnote{
As stated in Carrfs self-similarity hypothesis~\cite{B.Carr},
the collapsing spacetime near the singularity might evolve to a self-similar form. 
It is unclear, however, whether this hypothesis holds or not in the simplified model 
such as the dust collapse dealt in this paper. 
Thus, it would be fair to say that the assumption of the self-similarity in this paper is just for simplicity at this point.
}. 
This assumption corresponds to choosing 
	\begin{equation}\label{eq:F}
	 F(r)=\zeta r^n.
	\end{equation}
Here, $\zeta$ is a positive constant that we call the mass parameter. 
The total mass of the dust fluid is 
	\begin{equation}\label{eq:Total_mass}
	 2M=nF(r_0)=n\zeta r_0^n, 
	\end{equation}
where
$r = r_0$ represents the outer boundary of the fluid.
Note that the physical mass (ADM mass) defined in the asymptotic region is given by~\cite{Myers_1986}
	\begin{equation}\label{eq:ADM_mass}
	 M_{\mathrm{phys}}=\frac{(n+1)\Omega_{n+1}}{8\pi n}M,
	\end{equation}
where $\Omega_{n+1}=2\pi^{(n+2)/2}/\Gamma\left[(n+2)/2\right]$ is the volume of the unit $(n+1)$-sphere. Then, we find
	\begin{equation}\label{eq:r_0}
	 r_0=\left(\frac{16\pi M_{\mathrm{phys}}}{(n+1)\zeta \Omega_{n+1}}\right)^{1/n}.
	\end{equation}
Substituting Eqs.~(\ref{eq:tc}) and (\ref{eq:F}) into Eq.~(\ref{eq:t}),
we find 
	\begin{equation}
	 \tilde{R}(t,r)=r\left(1-\frac{n+2}{2}\sqrt{\zeta}\frac{t}{r}\right)^{2/(n+2)}.
	 \label{tilde_R}
	\end{equation}

At the boundary of the cloud ($r=r_0$), 
we match the LTB spacetime with the Schwarzschild-Tangherlini spacetime
	\begin{equation}\label{eq:S-T}
	 ds^2= -f(R) dT^2 + f(R)^{-1} dR^2
         +R^2d\Omega_{n+1}^2, 
\;\;\;
	f(R) := 1-\frac{2M}{nR^n}.
	\end{equation}
The metrics (\ref{eq:high-LTB}) and (\ref{eq:S-T}) describe the inside and the outside of the cloud, respectively. 
A matching condition is obtained by requiring that the metric of 
the $(n+2)$-dimensional boundary should be continuous. 
The results are 
\begin{align}
\begin{split}
	R(t)
	&= 
	\tilde{R}(t,r_0),
\\
	T(t)
	&= 
	- \sqrt{\frac{n}{2M}} \frac{2}{n+2} \tilde{R}^{(n+2)/2} (t,r_0)
		-\sqrt{\frac{n}{2M}} \int^{\tilde{R}(t,r_0)} \frac{ 2M R^{n/2}}{ nR^n - 2M } d R.
\label{juncCond}
\end{split}
\end{align}
The derivation of these matching conditions is given in~\ref{junction}.

\subsection{Double null coordinates}

In order to investigate the particle creation with the geometric-optics approximation, it is essential to solve the null geodesic equations or equivalently to obtain a double null coordinate system. In this paper, we adopt the latter way. In the rest of this section, we obtain the double-null form of the LTB metric and connect it to null coordinates in the outer region.

First, to obtain null coordinates inside the cloud
we introduce
\begin{equation}
	x := \frac{t}{r}, \;\;\; z := \ln r .
\end{equation}
Then, the metric (\ref{eq:high-LTB}) becomes
	\begin{equation}
	 ds^2=-r^2\left[x^2-\left(\partial_r \tilde{R}\right)^2\right]\left(d\tau^2-d\chi^2\right)+  \tilde{R}^2 d\Omega_{n+1}^2,
	\end{equation}
where
	\begin{equation}
	 \tau :=z+\frac{1}{2}\left(I_-+I_+\right),
\;\;\;
	\chi :=\frac{1}{2}\left(I_--I_+\right),
\;\;\;
	 I_\pm := \int^x \frac{dx}{x\pm \partial_r \tilde{R}}.
	\end{equation}
Then, a double-null coordinate system is given by
	\begin{equation}\label{eq:doubule_null0}
	 u=
         \left\{
         \begin{array}{cc}
	  &+re^{I_-} \quad \mathrm{for}\quad  x-\partial_r\tilde{R}>0 \\
          &-re^{I_-} \quad \mathrm{for}\quad x-\partial_r\tilde{R}<0
         \end{array}
	 \right.
         ,\quad 
         v=
         \left\{
         \begin{array}{cc}
	  &+re^{I_+} \quad\mathrm{for}\quad x+\partial_r\tilde{R}>0 \\
          &-re^{I_+} \quad\mathrm{for}\quad x+\partial_r\tilde{R}<0
         \end{array}
	 \right.,
        \end{equation}
where we have made these coordinates be identical to the standard Minkowski null coordinates in the flat limit $\zeta \rightarrow 0$.

To analyze a causal structure, we introduce a new variable
\begin{equation}\label{eq:def_y}
	y := \sqrt{ \frac{ \tilde{R}(t,r)} {r}  }.
\end{equation}
Note that $x$ and $y$ are related
\be
	x=\frac{2}{(n+2)\sqrt{\zeta}} \left( 1-y^{n+2} \right).
\ee

\begin{figure}[t]
\begin{center}
\includegraphics[height=5cm]{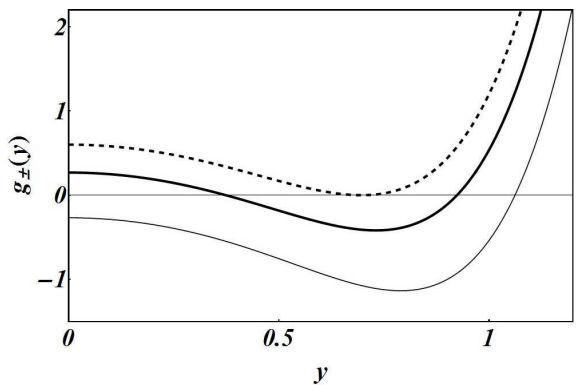}
\end{center}
\caption{{\it 
Numerical plots of $g_{\pm}(y)$ for the $n=2$ ($d=5$) case. The qualitative behaviors are independent of $n$. The thick solid curve is the $g_-(y)$ for $\zeta = 0.2 \times \zeta_c $; 
thick dashed curve is $g_-(y)$ for $\zeta = \zeta_c$; and the thin solid curve is $g_+(y)$ for $\zeta=0.2 \times \zeta_c $.
\label{g(y)}
}}
\end{figure}

\begin{figure}[bt]
\begin{center}
\setlength{\tabcolsep}{ 30 pt }
\begin{tabular}{ cc }
(a) & (b) \\
\includegraphics[height=5.5cm]{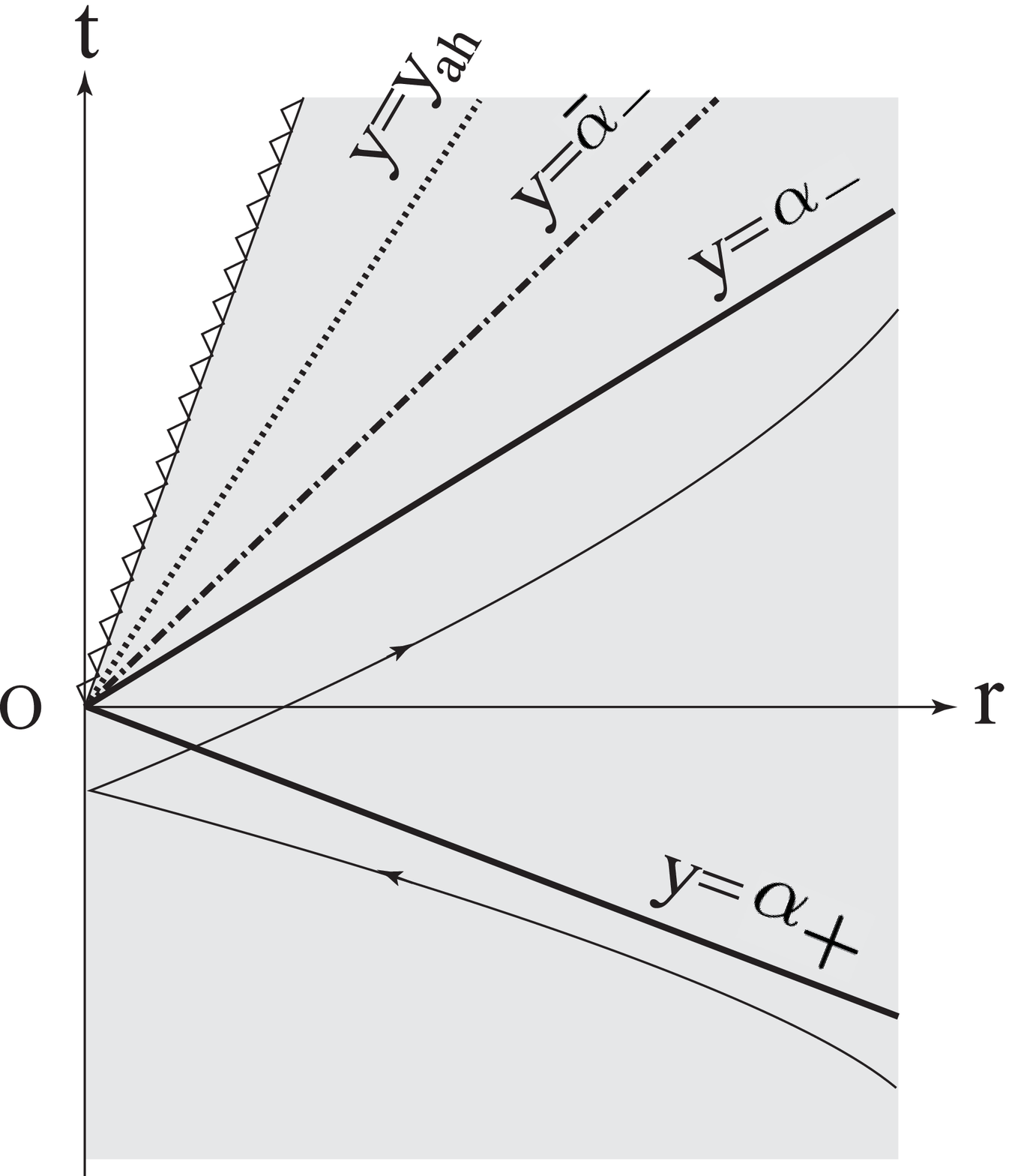} &
\includegraphics[height=5cm]{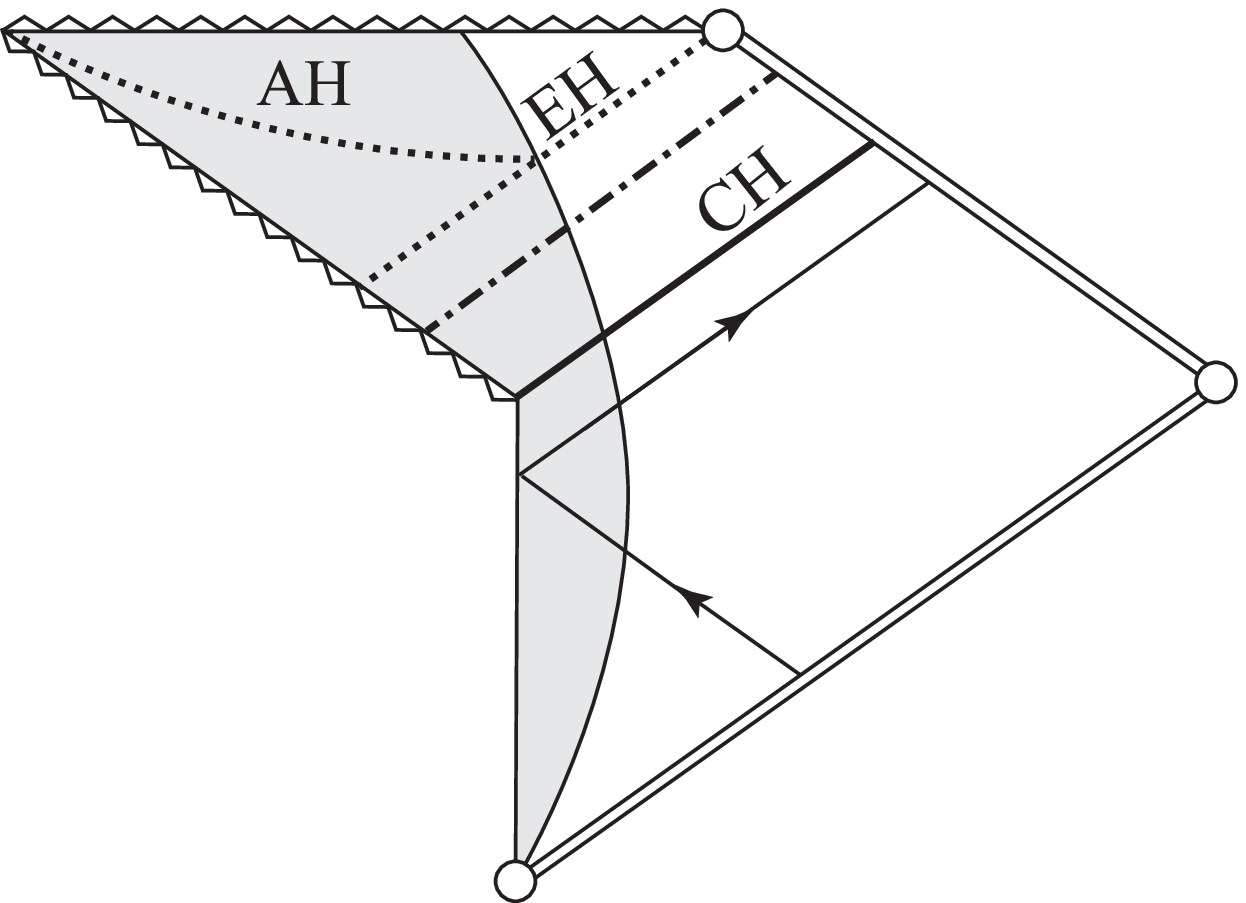} \\
\end{tabular}
\caption{{\it 
(a) A spacetime diagram in the $(t,r)$ coordinates. 
The Cauchy horizon $y=\alpha_-$ (thick solid), the null curve $y=\bar{\alpha}_-$ (dot-dashed), 
the apparent horizon $y=y_{\mathrm{ah}}$ (dotted), and the null curve which strikes the singularity $y=\alpha_+$
(thick solid) are drawn.
The curve with two arrows represents a typical null ray that passes through the regular center just before the 
appearance of singularity.
(b) A conformal diagram of the collapsing spacetime, that manifests the global nakedness of singularity. The Cauchy horizon (CH, thick solid), the null curve $y=\bar{\alpha}_-$ (dot-dashed), the apparent horizon (AH, dotted), and the event horizon (EH, dotted) are drawn. The shaded region is filled with the dust fluid infalling toward the center.
When $\zeta=\zeta_c$ (and in the limit $\zeta \rightarrow \zeta_c$), CH (thick solid) 
and $y=\bar{\alpha}_-$ (dot-dashed) merge. Note that even in this case the key structures do not change, i.e., the singularity is still globally naked.
\label{fig1}}}
\end{center}
\end{figure}

In terms of $y$, integrals $I_\pm$ can be written as 
	\begin{equation}\label{eq:int0}
	 I_\pm=2(n+2)\int^y \frac{y^{2n+1}dy}{g_{\pm}(y)},
\;\;\;
	 g_\pm(y) := 2y^{2n+2}\mp n\sqrt{\zeta}y^{n+2}-2y^n\mp2\sqrt{\zeta}.
	\end{equation}
Then, Eq.~(\ref{eq:doubule_null0}) becomes
	\begin{equation}\label{eq:doubule_null}
	 u=
         \left\{
         \begin{array}{cc}
	  &+re^{I_-} \quad \mathrm{for}\quad g_{-}(y)<0 \\
          &-re^{I_-} \quad \mathrm{for}\quad g_-(y)>0
         \end{array}
	 \right.
         ,\quad 
         v=
         \left\{
         \begin{array}{cc}
	  &+re^{I_+} \quad\mathrm{for}\quad g_+(y)<0 \\
          &-re^{I_+} \quad\mathrm{for}\quad g_+(y)>0
         \end{array}
	 \right..
        \end{equation}	
By simple algebra, we can see that $g_-(y)=0$ has two positive roots $\bar{\alpha}_{-}$ and $\alpha_{-}$ ($\bar{\alpha}_{-} \leq \alpha_{-}$) if and only if the mass parameter $\zeta$ is in the range $0<\zeta < \zeta_c$, where $\zeta_c$ is the critical value of the mass parameter. When $\zeta=\zeta_c$, two roots are degenerate $\bar{\alpha}_{-} = \alpha_{-} = \alpha_c$\footnote{For the $n=1$ case, this critical mass parameter is  $\zeta_c = [ 4(26-15\sqrt{3}) ]^{2/3} = 0.1809 \ldots $, as first obtained in \cite{Joshi:1993zg}. We present the explicit expressions of $\zeta_c$ and $\alpha_c$ for general $n$ in Appendix~\ref{app:zeta}}. The null ray $y=\alpha_{-}$ is the outgoing ray which emanates from the singularity ($u=v=0$), corresponding to the Cauchy horizon. We  can also find that $g_+(y)=0 $ has one  positive root $\alpha_+$. The null ray $y=\alpha_+$ is the incoming null ray which strikes the singularity ($u=v=0$). See Fig.~\ref{g(y)}, Fig.~\ref{fig1} and~\ref{ah}, which would help to understand the behaviors of $g_{\pm}(y)$ and causal structure.

Next, the null coordinates in the outer region are given by
	\begin{equation}\label{eq:Schnull}
	 U :=T- \int^R \frac{dR}{ f(R) },\quad V:=T + \int^R \frac{dR}{ f(R) },
	\end{equation}
with which the line element is given by
\begin{eqnarray}
	ds^2 = -f(R)dUdV + R^2 d\Omega_{n+1}^2.
\end{eqnarray}
Using the junction conditions \eqref{juncCond}, one can express ($U,V$) at the boundary ($r=r_0$) in terms of $y$,
\begin{align}
         U(y)&=-\sqrt{\frac{n}{2M}}\frac{2r_0^{(n+2)/2}y^{n+2}}{n+2}-r_0y^2-\int dy\frac{2r_0 y\left(2M+\sqrt{2nM}r_0^{n/2}y^n\right)}{nr_0^ny^{2n}-2M},  \label{eq:U}
\\
         V(y)&=-\sqrt{\frac{n}{2M}}\frac{2r_0^{(n+2)/2}y^{n+2}}{n+2}+r_0y^2+\int dy\frac{2r_0y\left(2M-\sqrt{2nM}r_0^{n/2}y^n\right)}{nr_0^ny^{2n}-2M}. \label{eq:V}
\end{align}

\section{Map of null rays}\label{sec:map_of_null}

In order to estimate the power and energy of the particle creation, it is necessary to obtain the map $V=G(U)$ between the retarded time $U$ where an incoming null ray departs the past null infinity and the advanced time $V$ where the null ray reaches the future null infinity after passing through the regular center $r=0$.
In this section, by the matching the inner null coordinates $u,v$ with the outer null coordinates $U,V$, we will obtain this map.

While it is difficult to obtain the general form of the map, one can easily obtain the map just before the Cauchy horizon as in the four-dimensional case~\cite{Brave_1998}. Thus, we will focus on the null rays in the neighborhood of $y=\alpha_{-}$ and $y=\alpha_c$. We investigate the generic case ($0<\zeta<\zeta_c$) and the degenerate case ($\zeta=\zeta_c$) separately.

\subsection{Generic case ($0<\zeta<\zeta_c$)}

For $0<\zeta<\zeta_c$,  we consider outgoing null rays near the Cauchy horizon $y\simeq \alpha_{-}$. In this case, Eq.~(\ref{eq:int0}) becomes
	\begin{equation}\label{eq:int01}
	 I_-=\gamma_-\ln\left(\frac{y-\alpha_{-}}{\alpha_{-}}\right),
\;\;\;
	 \gamma_-:=\frac{2(n+2) \alpha_-^{2n+1}}{g_-^{\prime}(\alpha_{-})}.
	\end{equation}
Substituting Eq.~(\ref{eq:int01}) into Eq.~(\ref{eq:doubule_null}) (the expression of $u$ for $ g_-(y) > 0 $), 
we find\footnote{Here, we set the integral region as $ \int_0^y $.}
	\begin{equation}\label{eq:y-a}
	 y-\alpha_{-}=\alpha_{-}\left(-\frac{u}{r}\right)^{1/\gamma_-}.
	\end{equation}
On the other hand, from Eq.~(\ref{eq:U}) we find that the
null ray near the Cauchy horizon on the boundary of the cloud ($r=r_0$) is given by\footnote{From Eq.~(\ref{eq:int0}) we obtain$ \sqrt{\zeta}=2\alpha_-^n(1-\alpha_-^{n+2})/(n\alpha_-^{n+2}+2) $.
Substituting this into $\alpha_-^n-\sqrt{\zeta}$, we find $\alpha_-^n-\sqrt{\zeta}=\alpha_-^{2(n+1)}(n+2)/(n\alpha_-^{n+2}+2)>0$, which ensures $ \Gamma_-<0 $.} 
\begin{equation}\label{eq:U1}
	 U(y) = U(\alpha_-) + \Gamma_- (y-\alpha_-),
\;\;\;
	 \Gamma_- := -\frac{2r_0 \alpha_-^{2n+1}}{\sqrt{\zeta}\left( \alpha_-^n-\sqrt{\zeta}\right)}<0.
	\end{equation}
Substituting Eq.~(\ref{eq:y-a}) into Eq.~(\ref{eq:U1}), we find the relation between $u$ and $U$ coordinates on the boundary of the cloud,
	\begin{equation}\label{eq:U2}
	 U(u)=U(\alpha_-)+\Gamma_-\alpha_- \left(-\frac{u}{r_0}\right)^{1/\gamma_-}.
	\end{equation}
This expression can be inverted to give 
	\begin{equation}\label{eq:u(U)}
	 u(U)=-r_0\left(-\frac{U_0-U}{\alpha_- \Gamma_-}\right)^{\gamma_-},
\;\;\;
	U_0 := U(\alpha_-).
	\end{equation}

Next, we focus on the incoming null rays near the curve $y\simeq \alpha_+$. 
For such a null ray, Eq.~(\ref{eq:int0}) becomes
	\begin{equation}\label{eq:int1}
	 I_+=\gamma_+\ln\left(\frac{y-\alpha_+}{\alpha_+}\right),
\;\;\;
	 \gamma_+ := \frac{ 2(n+2) \alpha_+ ^{2n+1} }{ g_+^\prime(\alpha_+) }.
	\end{equation}
Substituting Eq.~(\ref{eq:int1}) into Eq.~(\ref{eq:doubule_null}) (the expression of $v$ for $ g_+(y) > 0 $),  we find
	\begin{equation}\label{eq:y-a2}
	 y-\alpha_+=\alpha_+\left(-\frac{v}{r}\right)^{1/\gamma_+}.
	\end{equation}
From Eq.~(\ref{eq:V}), we find that the null ray near $y=\alpha_+$ on the boundary of the cloud ($r=r_0$) is given by 
	\begin{equation}\label{eq:V1}
	 V(y)=V(\alpha_+)+\Gamma_+(y-\alpha_+),
\;\;\;
	 \Gamma_+ :=
	-\frac{2r_0\alpha_+^{2n+1}}{\sqrt{\zeta}\left(\alpha_+^n+\sqrt{\zeta} \right)}<0.
	\end{equation}
Substituting Eq.~(\ref{eq:y-a2}) into Eq.~(\ref{eq:V1}), 
we find 
the relation between $v$ and $V$ coordinates on the boundary of the cloud,
	\begin{equation}\label{eq:V2}
	 V(v)=V_0+\Gamma_+\alpha_+\left(-\frac{v}{r_0}\right)^{1/\gamma_+},
\;\;\;
	V_0 := V(\alpha_+).
	\end{equation}

From Eqs.~(\ref{eq:u(U)}) and (\ref{eq:V2}), we obtain the desired map of null rays
$V=G(U)$ as
	\begin{equation}\label{eq:g0}
	 G(U) = V_0+\Gamma_+\alpha_+\left(-\frac{U_0-U}{\alpha_- \Gamma_-}\right)^{\gamma},
\;\;\;
	\gamma := \frac{\gamma_-}{\gamma_+}.
	\end{equation}
Note that we have put $u=v$ at the center ($r=0$), because the null ray is reflected at the center\footnote{
While it is possible to choose $u\neq v$ at the center, 
we take $u=v$ just according to Ref.~\cite{Brave_1998}.}.

\subsection{Degenerate case ($\zeta=\zeta_c$)}

We consider the outgoing null ray passing near the Cauchy horizon $y\simeq \alpha_c$ for $\zeta=\zeta_c$. In this case, Eq.~(\ref{eq:int0}) becomes
	\begin{equation}\label{eq:int2}
	 I_-=\gamma_c\left(-\frac{1}{y-\alpha_c}+\frac{1}{\alpha_c}\right),
\;\;\;
	 \gamma_c := \frac{4(n+2)\alpha_c^{2n+1}}{g_-^{\prime\prime}(\alpha_c)}.
	\end{equation}
Substituting Eq.~(\ref{eq:int2}) into Eq.~(\ref{eq:doubule_null}) (the expression for $g_-(y)>0$), 
we find
\begin{equation}\label{eq:y-a3}
	 y-\alpha_c=-\frac{\gamma_c}{\ln\left[A\left(-u/r\right)\right]},
\;\;\;
	A := \exp\left(-\gamma_c/\alpha_c\right).
\end{equation}
On the other hand, from Eq.~(\ref{eq:U}) we find that the null ray near the Cauchy horizon on the boundary of the cloud ($r=r_0$) is given by
	\begin{equation}\label{eq:U3}
	 U(y)=U(\alpha_c)+\Gamma_c (y-\alpha_c),
\;\;\;
	 \Gamma_c
	:= -\frac{2r_0 \alpha_c^{2n+1}}{\sqrt{\zeta_c}\left(\alpha_c^n-\sqrt{\zeta_c}\right)}<0.
\end{equation}
Substituting Eq.~(\ref{eq:y-a3}) into Eq.~(\ref{eq:U3})
we find the relation between $u$ and $U$ coordinates on the boundary of the cloud,
	\begin{equation}\label{eq:U4}
	 U(u)=U(\alpha_c)-\Gamma_c\frac{\gamma_c}{\ln\left[A\left(-u/r\right)\right]}.
	\end{equation}
This can be inverted to give 
	\begin{equation}\label{eq:u(U)2}
	 u(U)=-\frac{r_0}{A}\exp\left(\frac{\gamma_c\Gamma_c}{\tilde{U}_0-U}\right),
\;\;\;
	\tilde{U}_0 := U(\alpha_c).
	\end{equation}
From Eqs.~(\ref{eq:V2}) and (\ref{eq:u(U)2}), we obtain the desired map of null rays 
$V=G(U)$ as 
	\begin{equation}\label{eq:g2}
	 G(U)=V_0+\frac{\Gamma_+\alpha_+}{A^{1/\gamma_+}}\exp\left(\frac{\gamma_c\Gamma_c}{\gamma_+\left(\tilde{U}_0-U\right)}\right),
	\end{equation}
where we have put $u=v$ at the center again.

\section{Power and energy of particle emission}\label{sec:power_and_energy}

We evaluate the power and energy of the particle creation within the geometric-optics approximation. For simplicity, we focus on a massless scalar field $\phi$ coupled to the Ricci scalar curvature $\mathcal{R}$
as
	\begin{equation}
	 \left(\Box -\xi \mathcal{R} \right)\phi=0,
	\end{equation}
where $\xi$ is an arbitrary constant.
In particular, the cases of $\xi=0$ and $\xi = (n+1)/[4(n+2)]$ are 
called the minimal coupling and the conformal coupling, respectively.

The formula of the power for the general value of $\xi$ in general dimension $d$ was obtained in Ref.~\cite{Miyamoto_Nemoto_2011a}
according to~\cite{Ford:1978ip}
\footnote{
We present a notable comment which has less to do with the body of this paper as follows:
If one wants to estimate the spectrum, 
one has to define the out-vacuum in addition to the in-vacuum, 
and calculate the Bogoliubov coefficients. 
However, in our paper, we only estimate the expectation value of the energy-momentum tensor 
in the in-vacuum, which does not need the information about the out-vacuum. 
An almost complete formulation was presented in Appendix B of Ref.~\cite{Miyamoto_Nemoto_2011a}, 
which is just a higher-dimensional generalization of the original 
formulation given by Ford and Parker~\cite{Ford:1978ip}.
},
	\begin{equation}\label{eq:power}
	 P(U)=\frac{1}{4\pi}
	\left[
		\left(\frac{1}{4}-\xi\right)\left(\frac{G''(U)}{G'(U)}\right)^2
         +\left(\xi-\frac{1}{6}\right)\frac{G'''(U)}{G'(U)}
	\right].
	\end{equation}
The total energy radiated can be estimated by integrating the power over time $U$,
	\begin{equation}\label{eq:energy}
	 E(U)=\int_{-\infty }^{U}P(U)dU.
	\end{equation}

\subsection{Generic case ($0<\zeta<\zeta_c$)}

\begin{figure}[bt]
\begin{center}
\includegraphics[height=6cm]{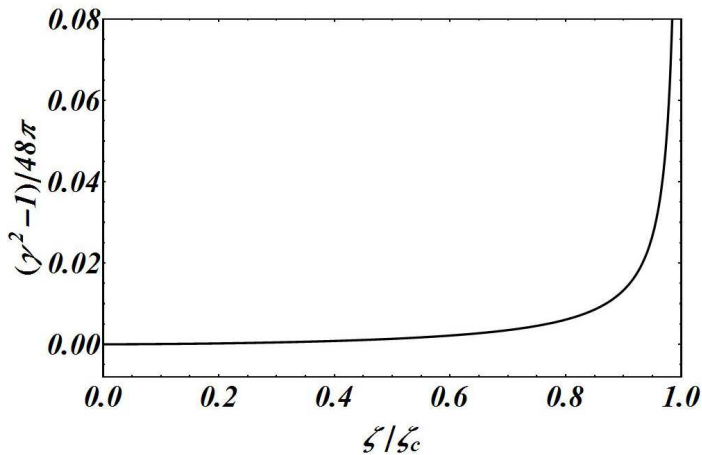}
\end{center}
\caption{{\it
$\zeta$-dependence of $\mathcal{F}$ for the minimally coupling scalar field ($\xi=0$) 
in $d=n+3=5$. Qualitative behaviors are independent of $d$ and $\xi$. 
\label{gamma^2-1}}}
\end{figure}

Substituting Eq.~(\ref{eq:g0}) into Eqs.~(\ref{eq:power}) and (\ref{eq:energy}), 
we find
	\begin{align}
	\begin{split}\label{eq:generic_result}
	 P(U)&=\mathcal{F}\left(U_0-U\right)^{-2},
	\\
	 E(U)&=\mathcal{F}\left(U_0-U\right)^{-1},
	 \end{split}
        \end{align}
where
	\begin{align}\label{eq:factor_energy_power}
	 \mathcal{F}:=\frac{(\gamma-1)(\gamma+1-12\xi)}{48\pi}.
	\end{align}
Thus, we reproduce and generalize to general dimensions the result in Ref.~\cite{Brave_1998}. The power diverges as the quadratic inverse of the remaining time to the Cauchy horizon.
The $\zeta$-dependence of $\mathcal{F}$ for the minimally coupling scalar field ($\xi=0$) in $d=n+3=5$ is shown in Fig.~\ref{gamma^2-1}. The factor $\mathcal{F}$ closes to zero in the limit $\zeta \rightarrow 0$, and increases monotonically with $\zeta$ to diverge in the limit $\zeta\rightarrow \zeta_c$. Qualitative behaviors are independent of $d$ and $\xi$.

\subsection{Degenerate case ($\zeta=\zeta_c$)}

Substituting Eq.~(\ref{eq:g2}) into Eqs.~(\ref{eq:power}) and (\ref{eq:energy}), 
we find
	\begin{align}
	\begin{split}\label{eq:degenerate_P_E0}
	 P(U)&=\frac{\gamma_c^2\Gamma_c^2}{48\gamma_+^2}\left(\tilde{U}_0-U\right)^{-4},\\
	 E(U)&=\frac{\gamma_c^2\Gamma_c^2}{144\gamma_+^2}\left(\tilde{U}_0-U\right)^{-3}.
		\end{split}
        \end{align}
Substituting Eqs.~(\ref{eq:r_0}) and (\ref{eq:U3}) into Eq.~(\ref{eq:degenerate_P_E0}),
we find
	\begin{align}
	\begin{split}\label{eq:degenerate_P_E}
	 P(U)&=\frac{E_P}{t_P}\mathcal{F}_c\left(\frac{t_P}{\tilde{U}_0-U}\right)^{4},
	\\
	 E(U)&=E_P\mathcal{F}_c\left(\frac{t_P}{\tilde{U}_0-U}\right)^{3},
        \end{split}
        \end{align}
where 
	\be\label{eq:order_of_power}
	 \mathcal{F}_c := \left(\frac{M_{\mathrm{phys}}}{M_P}\right)
         \left(\frac{16\pi }{(n+1)\zeta_c\Omega_{n+1}}\right)^{2/n}\frac{\gamma_c^2\alpha_c^{2(2n+1)}}{12\gamma_+^2\zeta_c \left(\alpha_c^n-\sqrt{ \zeta_c }\right)^2}.
	\ee
In the final expressions (\ref{eq:degenerate_P_E}) and (\ref{eq:order_of_power}), we have revived the natural constants. The Planck time $t_P $, mass $M_P$, and energy $E_P$ are given by
\be
	t_P
	:=
	\left( \frac{ \hbar G }{ c^{d+1} } \right)^{1/(d-2)},
\;\;\;
	M_P
	:=
	\frac{E_P}{c^2}
	:=
	\left( \frac{ \hbar^{d-3} }{ c^{d-5}G  } \right)^{1/(d-2)}.
\ee

The power and energy~(\ref{eq:degenerate_P_E}) diverge as the quartic and cubic inverse of the remaining 
time to the Cauchy horizon, respectively.
Note that the power and energy depend on the total mass of the dust fluid $M_{\mathrm{phys}}$ 
in spite of the scale invariance of the central self-similar region.
This result is in contrast to the result in the limit $\zeta\rightarrow \zeta_c$ of the generic case (\ref{eq:generic_result}). We also mention that the coupling constant $\xi$ does not appear in Eq.~(\ref{eq:degenerate_P_E}). These features are similar to those in the higher-dimensional Vaidya (null dust) case~\cite{Miyamoto_Nemoto_2011a}. Numerical values of \eqref{eq:order_of_power} when the parameters are chosen as $M_{P}c^2=1 \; \mathrm{TeV}$ and 
$M_{\mathrm{phys}}c^2=5 \; \mathrm{TeV}$ are given in Table \ref{tab:example}.

\begin{table}[bt]
\caption{{\it 
Numerical values of $\zeta_c$, $\alpha_c$, and $\mathcal{F}_c$ in Eq.~\eqref{eq:order_of_power} when $M_{P} c^2=1 \; \mathrm{TeV}$ and $M_{\mathrm{phys}}c^2=5 \; \mathrm{TeV}$.
\label{tab:example}
}}
\footnotesize\rm
\begin{tabular*}{\textwidth}{@{}l*{15}{@{\extracolsep{0pt plus12pt}}l}}
\br
$d=n+3$ & 4 & 5 & 6 & 7 & 8 & 9 & 10 \\
\mr
$\zeta_c$  &  0.181 & 0.0902 & 0.0564 &0.0394 &0.0294 & 0.0230 & 0.0185 \\
$\alpha_c$  &  0.581 & 0.697 & 0.756 & 0.794 & 0.820 & 0.840 & 0.854 \\
$\mathcal{F}_c$ & 824 & 0.646 & 0.210 & 0.128 & 0.0957 & 0.0817 & 0.0743 \\
\br
\end{tabular*}
\end{table}

\section{Conclusion and discussions}\label{sec:conclusion}

Not only mini black holes but also naked singularities can be
produced by high-energy particle collisions in the context of TeV-scale gravity.  
Therefore, we may detect the sign of extra dimensions in collider experiments. In such a context, we have estimated the power and energy of the particle creation by the naked singularity 
in the $d$-dimensional collapsing spacetime. The background has been modeled by the spherical collapse of 
inhomogeneous dust solution, called the LTB solution~(\ref{eq:high-LTB}).

We have obtained the formulae of the emission power and energy Eq.~(\ref{eq:generic_result}) for the generic case~($0<\zeta<\zeta_c$) and Eq.~(\ref{eq:degenerate_P_E}) for the degenerate case~($\zeta=\zeta_c$),
where $\zeta$ is the mass parameter~(\ref{eq:F}).
In the generic case, the power and energy diverge as the quadratic and 
linear inverse of the remaining time to the Cauchy horizon, respectively, 
and have reproduced and generalized the four-dimensional result.
In the degenerate case, we have found that the power and energy diverge as the quartic and cubic 
inverse of the remaining time, respectively, 
and have shown that the power and energy depend on the total mass of dust fluid $M_{\mathrm{phys}}$ 
despite the scale invariance of the central region.
While a similar result has been obtained in the Vaidya case~\cite{Miyamoto_Nemoto_2011a}, 
it should be stressed that the behavior of the power and energy is different between the exactly degenerate case $\zeta=\zeta_c$ and 
the limit of $\zeta\rightarrow \zeta_c$.

We have adopted the geometric-optics approximation to estimate the power and energy.
This approximation is justified if the wavelength of the mode function is shorter enough than the curvature radius of 
the background spacetime, in particular, near the central region. 
The fact that the redshift diverges at the Cauchy horizon 
(see \ref{app:redshift}) 
seems to support the validity of the geometric-optics approximation. 
We have to be careful, however, since the background curvature can be arbitrarily large near the singularity. 
One way to estimate the size of waves involved in the particle creation is to trace the wave at the future null infinity backward in time until the center. 
As mentioned in the introduction, however, to estimate the wavelength at the future null infinity (i.e., to estimate the spectrum), 
one has to know the map $V=G(U)$ beyond the Cauchy horizon, namely, the boundary condition at the singularity is required. 
Therefore it seems difficult to prove or disprove the validity of the geometric-optics approximation systematically at this point.

Recently, M. Banados, J. Silk and S. M. West considered collision of two particles near the event horizon of a Kerr spacetime 
and showed that the center-of-mass energy diverges (BSW effect)~\cite{BSW}. 
This result has been extended to the collision near a naked singularity. For example, in the Reissner- Nordstrom naked singularity 
it was shown that the center-of-mass energy related to two colliding particles is divergent~\cite{Patil and Joshi et.al}.
Therefore, it is possible that high-energy particles associated with the BSW effect near naked singularities reach 
the asymptotic region as the quantum particles considered in this paper. It will be interesting to compare the two phenomena 
near the naked singularities, i.e., the BSW effect and the particle creation, from the viewpoints of emission power and/or total energy.

It is quite interesting to investigate the actual visibility (or observability) of the naked singularities, 
for example, by calculating the spectrum and redshift of created particles (see Refs.~\cite{M. Bejger} and ~\cite{Z. k. Stuchlik}
for the related investigation for the BSW effects).
However, the calculation of spectrum (i.e., the Bogoliubov coefficients) 
faces with the fundamental problem that the boundary condition of the quantum field at the naked singularity cannot be determined a priori. 
Therefore, we have no clear answer to the visibility of the naked singularity at this point unfortunately.

Finally, we should stress that the quantum effect investigated in the present TeV-scale gravity context will affect the following time evolution of the spacetime, in contrast to the case of stellar collapse. 
Namely, as argued in Harada {\it et al}.~\cite{Harada:2000me}, if any quantum gravitational effect works as the cutoff of the radiation, 
the net energy radiated only amounts to the Planck energy. 
This suggests that a backreaction is negligible in the collapse of stellar-size massive objects. 
On the other hand, in the present TeV-scale gravity context, both the energy of background (determined by the energy scale of the collider experiment) and the radiated energy can be the order of TeV. 
Thus, the background can be affected strongly by the quantum radiation. 
It would be interesting to investigate the semiclassical dynamics of spacetime and any quantum gravitational effects in the view of this possibility.

\section*{Acknowledgments}

We would like to thank H.~Kodama and T.~Harada for useful comments and discussions. 
This work is supported by the JSPS Grant-in-Aid for Young Scientists (B) No.\ 25800157.

\appendix

\section{Comments on relations among preceding papers and this paper}
\label{relation}

In this section, we discuss the consistency of the argument of the visible-border 
production~\cite{Nakao_2010}, by which this paper is motivated, 
with that in the earlier well-known papers such as Giudice et al.~\cite{Giudice:2002} 
and Giddings-Rychkov~\cite{Giddings:2004}, 
and mention the relation 
between Ref.~\cite{Nakao_2010} and the present paper.

\subsection{Nakao et al.~\cite{Nakao_2010} and Giudice et al.~\cite{Giudice:2002}}

While Nakao et al.~\cite{Nakao_2010} claims that the trans-Planckian collision in the TeV-scale gravity may lead to the production of visible borders of spacetime, the well-known paper by Giudice et al.~\cite{Giudice:2002} shows the trans-Planckian collisions are semi-classical. Let us argue below that this is just an apparent inconsistency.

In Ref.~\cite{Giudice:2002} the authors focus on the trans-Planckian regime, where the center-of-mass energy of colliding particles is greater than the fundamental Planck energy. As they argued, in the trans-Planckian regime the Schwarzschild radius of the system is much greater than the fundamental Planck scale and the de Broglie wavelength of the colliding particles, and therefore characterizes the dynamics. From this reason, the trans-Planckian regime corresponds to a classical limit but the magnitude of curvature around the collision is not restricted in this regime.  On the other hand, Ref.~\cite{Nakao_2010} supposes that the colliding particles have an impact parameter larger than the Schwarzschild radius but the center-of-mass energy is trans-Planckian. Once this situation is realized, the magnitude of curvature around the colliding region can become trans-Planckian but the horizon will not form. Therefore, even in the trans-Planckian regime the trans-Planckian curvature region is visible, i.e., the visible borders of spacetime can appear. 

Moreover, in the present paper, to obtain the energy and power of particle creation in the higher-dimensional LTB spacetime we did not apply the quantum gravitational analysis but applied the semi-classical analysis to this spacetime. Therefore, the possible existence of visible borders of spacetime and our analysis do not contradict the argument in Ref.~\cite{Giudice:2002}.

\subsection{Nakao et al.~\cite{Nakao_2010} and Giddings-Rychkov~\cite{Giddings:2004}}

While Ref.~\cite{Nakao_2010} claims that the trans-Planckian collision may lead to the effective naked singularities, the well-known paper~\cite{Giddings:2004} shows that the curvature remains finite once quantum uncertainties are properly taken into account. 

Giddings and Rychkov~\cite{Giddings:2004} consider the collision of two ultra-relativistic point particles using the Aichelburg-Sexlmetric in the trans-Planckian regime, and show that small curvature can be obtained even at the instant of intersection of particles by choosing special size of wavepackets. On the other hand, Ref.~\cite{Nakao_2010} considers the {\it arbitrary} size of wavepacket with specifying {\it no} background geometry, and shows that the visible borders can appear. Therefore, the consideration in~\cite{Nakao_2010} is more general in a sense than that in \cite{Giddings:2004}. It is added that in the modeling of~\cite{Nakao_2010} even if quantum uncertainties were taken into account, the trans-Planckian region of spacetime can appear unless choosing size of wavepackets. 

\subsection{Nakao et al.~\cite{Nakao_2010} and the present paper}

Our paper is certainly motivated by the paper~\cite{Nakao_2010}, 
that argues the appearance of visible borders of spacetime. 
However, the background spacetime assumed in our paper just 
represents the naked-singularity formation in higher-dimensional 
spherical collapse and therefore, in a practical sense, has nothing 
to do with the visible border that could be generated in collider experiments. 
In addition, the particle creation is a quite fundamental phenomenon that can 
play important roles in various situations. Therefore, we believe that our analyses and general results presented in fully analytic form have their own value independently of the motivation.

\section{Junction conditions at the dust-vacuum boundary}
\label{junction}

The boundary $\Sigma$ between the LTB and Schwarzschild regions can be parameterized as $ (T,R)=(T(t),R(t)) $, where $t$ is the time coordinate in the LTB region. The induced metric (or the first fundamental form) on $\Sigma$ is given in both coordinate patches as
\be
&&
	ds_\Sigma^2
	=
	-dt^2 + \tilde{R}^2 (t,r_0) d\Omega_{n+1}^2,
\\
&&
	ds_\Sigma^2
	=
	\left[
		- f(R) \left( \frac{ dT }{ dt } \right)^2
		+ f(R)^{-1} \left( \frac{ dR }{ dt } \right)^2
	\right] dt^2 + R^2 d\Omega_{n+1}^2.
\ee
Equating the above expressions, we obtain first junction conditions
\be
&&
	R(t)=\tilde{R}(t,r_0),
\label{junc1}
\\
&&
	- f(R) \left( \frac{ dT }{ dt } \right)^2
	+ f(R)^{-1} \left( \frac{ dR }{ dt } \right)^2 = -1.
\label{junc2}
\ee
Using Eq.~\eqref{junc1} and the Einstein equation \eqref{eq:dotR}, we obtain
\be
	\frac{ dR }{ dt } = \frac{ \pd \tilde{R} }{ \pd t } = \sqrt{ \frac{ F(r_0) }{ R^n } },
\;\;\;
	\frac{ dT }{ dt }
	=
	\frac{ dR }{ dt } \frac{ dT }{ dR }
	=
	\sqrt{ \frac{ F(r_0) }{ R^n } } \frac{ dT }{ dR }.
\label{junc3}
\ee
Plugging these relations into \eqref{junc2}, we obtain
\be
	T(R)
	&=&
	-\frac{1}{\sqrt{F(r_0)}} \int^R \frac{ R^{n/2} }{ f(R) } dR.
\ee
Using the explicit expression of $f(R)$, one can derive the second equation in \eqref{juncCond}.

The extrinsic curvature (or the second fundamental form) on $\Sigma$ is defined by $K_{ab}=\nabla_\mu n_\nu e^\mu_a e^\nu_b$,
where $n_\mu$ and $e^\mu_a$ (a,b=1,2,\ldots,n) are the unit normal and tangent vectors to $\Sigma$.
The unit normal to $\Sigma$ is given in both coordinates as 
	\be
	&&
	 n_\mu=\left(\partial_r\tilde{R}\right)dr_\mu,
	\\
	&&
	 n_\mu=-\left( \frac{ dR }{ dt } \right)dT_\mu+\left( \frac{ dT }{ dt } \right)dR_\mu,
        \ee
where these expressions are obtained from the relations $n_\mu e^\mu_a=0$, $n_\mu n^\mu=1$.
From a straightforward calculation, we obtain the extrinsic curvature on both sides of $\Sigma$ as
	\be
	&&
	 K_{11}=0, \quad K_{22}=K_{33}=\cdots=K_{nn}=\left[\tilde{R}\left(\partial_r\tilde{R}\right)\right]^{-1},\label{eq:extrinsic_1}
	\\
	&&
	 K_{11}=\left(\frac{d\epsilon}{dt}\right)\left( \frac{ dR }{ dt } \right)^{-1}, 
         \quad K_{22}=K_{33}=\cdots=K_{nn}=\epsilon R^{-1},\label{eq:extrinsic_2}
        \ee
where
	\be
         \epsilon=f\left( \frac{ dT }{ dt } \right)=\sqrt{\left( \frac{ dR }{ dt } \right)^2+f}.
        \ee
Equating Eqs.~(\ref{eq:extrinsic_1}) and~(\ref{eq:extrinsic_2}), we obtain second junction conditions
	\be
        &&
        \left(\frac{d\epsilon}{dt}\right)\left( \frac{ dR }{ dt } \right)^{-1}=0,
        \\
        &&
         \left[\tilde{R}\left(\partial_r\tilde{R}\right)\right]^{-1}=\epsilon R^{-1}.
        \ee

\section{Explicit expressions of $\zeta_c$ and $\alpha_c$}\label{app:zeta}

Let us obtain the relation between the mass parameter $\zeta$ and the root $\alpha$.
From $g_\pm(y)=0$, we can express the mass parameter $\zeta$ in two ways,
	\be\label{eq:azeta}
	 \sqrt{\zeta}=\frac{2\alpha_+^n\left(\alpha_+^{n+2}-1\right)}{n\alpha_+^{n+2}+2}=\frac{2\alpha_-^n\left(1-\alpha_-^{n+2}\right)}{n\alpha_-^{n+2}+2}.
	\ee
By using Eq.~(\ref{eq:azeta}), $\gamma$ in Eq.~(\ref{eq:g0}) is given by 
	\be
         \gamma=\frac{2n^2\alpha_+^{n+2}+4(3n+2)-4n\alpha_+^{-(n+2)}}{2n^2\alpha_-^{n+2}+4(3n+2)-4n\alpha_-^{-(n+2)}}.
        \ee
Thus, we obtain
	\be\label{eq:agamma}
         \gamma-1=\frac{2n^2\left(\alpha_+^{n+2}-\alpha_-^{n+2}\right)+4n\left(\alpha_+\alpha_-\right)^{-(n+2)}\left(\alpha_+^{n+2}-\alpha_-^{n+2}\right)}{2n^2\alpha_-^{n+2}+4(3n+2)-4n\alpha_-^{-(n+2)}}>0,
        \ee
where we have used $\alpha_+>\alpha_-$, 
and have checked numerically that the denominator of Eq.~(\ref{eq:agamma}) 
is positive over a wide range of dimensions ($1 \leq n \leq 7$).

The critical value of the mass parameter $\zeta_c$ can be expressed by $\alpha_c$. 
From $g_-(y)=g'_-(y)=0$, we find 
	\be\label{eq:azetacrit}
         \sqrt{\zeta_c}=\frac{2\alpha_c^n\left(1-\alpha_c^{n+2}\right)}{n\alpha_c^{n+2}+2}=\frac{2\left[n-2(n+1)\alpha_c^{n+2}\right]}{n(n+2)\alpha_c^2}.
        \ee
Then, we find the critical value of $\alpha$ in terms of $n$,
	\be
         \alpha_c=\left[(n+2)\sqrt{2n+1}-3n-2\right]^{1/(n+2)}n^{-2/(n+2)}.
        \ee

\section{The position of the singularity and apparent horizon}
\label{ah}

In order to understand the spacetime structure near the singularity, 
we have to find the position of the spacetime singularity and apparent horizon in addition to the Cauchy horizon. 

The singularity curve \eqref{eq:t} for the present choice of $F(r)$, equation \eqref{eq:F}, is equivalently given by 
\be
	x=
	\frac{ 2 }{ (n+1) \sqrt{\zeta} }.
\ee

The apparent horizon is a $(n+1)$-sphere on the $(n+2)$-spacelike hypersurface which satisfies the following condition
	\be
	 \theta_-<0,\quad \theta_+=0,
	\ee
where $\theta_\pm$ is the null expansions~\cite{Hawking}
	\be\label{eq:expansion}
 	 \theta_{\pm}
	=
	\nabla_\mu N_\nu^{(\pm)}\left(\sum^{n+1}_{i=1} Y_{(i)}^\mu Y_{(i)}^{\nu}\right).
	\ee
Here, $Y^\mu_{(i)}$ denotes the tangent vectors for the $(n+1)$-spacelike hypersurfaces.
In our study, we focus on the vectors tangent to $(n+1)$-sphere as follows
	\begin{align} \nn
		Y_{(1)}^\mu
		=
		\frac{1}{\tilde{R}}\left(\frac{ \pd }{ \pd \theta_1}\right)^\mu, &\;\;\;
         Y_{(2)}^\mu =\frac{1}{\tilde{R}\sin\theta_1}\left(\frac{ \pd }{ \pd \theta_2 }\right)^\mu,
		\\
                \ldots,&\;\;\;
         Y_{(n+1)}^\mu =\frac{1}{\tilde{R}\sin\theta_1\cdots\sin\theta_n}\left(\frac{ \pd }{\pd \theta_{n+1}}\right)^\mu.
         \label{eq:tangent_sphere}
         \end{align}
$N_\mu^{\pm}$ is the normal vectors for the $(n+1)$-sphere,
	\be\label{eq:normal_sphere}
 	 N_\mu^+=(dt)_\mu+\partial_r\tilde{R}(dr)_\mu,\quad 
         N_\mu^-=(dt)_\mu-\partial_r\tilde{R}(dr)_\mu.
	\ee
Substituting Eqs.~(\ref{eq:tangent_sphere}) and (\ref{eq:normal_sphere}) into 
Eq.~(\ref{eq:expansion}), we obtain 
	\be
	 \theta_- = \frac{n+1}{\tilde{R}}\left(\partial_t\tilde{R}-1\right),
	\;\;\;
       	 \theta_+ = \frac{n+1}{\tilde{R}}\left(\partial_t\tilde{R}+1\right).
	\ee
Because we have demanded $\partial_t\tilde{R}<0$ to obtain the collapse solution, we get $\theta_-<0$.
To find the apparent horizon we solve $\theta_+=0$, i.e., $\partial_t\tilde{R}=-1$ with respect to $y$.
By using Eqs.~(\ref{eq:dotR}),~(\ref{eq:F}),~(\ref{tilde_R}), and~(\ref{eq:def_y}), we find the apparent horizon $y_{\rm ah}$,
	\be
	 y_{\rm ah}=\zeta^{1/2n}.
	\ee

We can find the apparent horizon also from the following fact:
the apparent horizon is identified with the curve where the circumferential radius $\tilde{R}$ does not increase along an outgoing null ray. 
Since $dt/dr=\pd_r \tilde{R}$ along the null ray from equation \eqref{eq:high-LTB}, 
the derivative of $\tilde{R}$ by $r$ along the null ray $t=t(r)$ is
\be
	\frac{ d\tilde{R} (t(r),r) }{dr}
	=
	\frac{ \pd \tilde{R} }{ \pd t } \frac{ dt }{ dr }  + \frac{ \pd \tilde{R} }{ \pd r }
	=
	\frac{ \pd \tilde{R} }{ \pd r }
	\left(
		\frac{ \pd \tilde{R} }{ \pd t } + 1
	\right).
\ee
Using the explicit form of $\tilde{R}$ in \eqref{tilde_R} we finally obtain
\be
	\frac{ d\tilde{R} (t(r),r) }{dr} = \frac{ ny^{n+2}+2 }{ (n+2)y^n }
	\left(
		1-\frac{ \sqrt{\zeta} }{ y^n }
	\right).
\ee
Thus, the apparent horizon, where $ d\tilde{R}/dr = 0 $, turns out to be
\be
	y = y_{\rm ah} = \zeta^{1/2n}
\;\;\;
\mbox{or equivalently}
\;\;\;
	x=
	\frac{2}{(n+2) \sqrt{\zeta}} \left( 1-\zeta^{(n+2)/2n} \right).
\ee

\section{Gravitational redshift of null rays}\label{app:redshift}

Let us estimate the redshift of the radial null rays in the generic and degenerate cases. 
The tangent of a null geodesic $k^\mu=dx^\mu/d\lambda$, where $\lambda$ is an affine parameter, 
is obtained by solving $k^\mu \nabla_\mu k^\nu=0$.
The $t$-component of this equation is 
	\be\label{eq:null_geodesic}
         \frac{dk^t}{d\lambda}+\frac{\partial_t\partial_r\tilde{R}}{\partial_r\tilde{R}}\left(k^t\right)^2=0.
        \ee
From a null condition $k^\mu k_\mu=0$, we find
	\be
         k^r=\frac{1}{\partial_r\tilde{R}}k^t.
        \ee
Using this condition, we obtain the following relation for the derivative of a function of $y$,
	\be\label{eq:relation_y}
         \frac{d}{d\lambda}=-\frac{g_\pm(y)}{2ry^{n+1}\left(ny^{n+2}+2\right)}k^t\frac{d}{dy}.
        \ee
Substituting Eq.~(\ref{eq:relation_y}) into Eq.~(\ref{eq:null_geodesic}), we can rewrite Eq.~(\ref{eq:null_geodesic}) as 
	\be\label{eq:nullcondequation}
         \frac{dk^t}{dy}-K_\pm(y)k^t=0,
        \ee
where
	\be
         K_\pm(y)=\frac{2(n+2)y^{2n+1}}{g_\pm(y)}\left[1+\frac{g'_{\pm}(y)}{2(n+2)y^{2n+1}}-\frac{ng_\pm(y)}{2(n+2)y^{2(n+1)}}\right].
        \ee
A formal solution of Eq.~(\ref{eq:nullcondequation}) is 
	\be\label{eq:geodesic_equation}
         \frac{k^t(y)}{k_0^{t,\pm}}=\exp\left[\int K_\pm(y) dy \right],
        \ee
where $k_0^{t,\pm}$ is an integration constant.

\subsection{Generic case ($0<\zeta<\zeta_c$)}

In the generic case, $g_-(y)=0$ has the non-degenerate root at $y=\alpha_-$ and $y=\bar{\alpha}_-$, 
and $g_+(y)=0$ has the root $y=\alpha_+$.
Subtracting these poles of the integrand in Eq.~(\ref{eq:geodesic_equation}), we have
	\begin{align}\label{eq:redshift_generic}
         \nonumber \frac{k^t(y)}{k_0^{t,\pm}}&=\exp\left[-\int_{\tilde{y}_0^\pm}^y
         \frac{2(n+2)\alpha_\pm^{2n+1}}{g_\pm'(\alpha_\pm)\left(\tilde{y}-\alpha_\pm\right)}\left[1+\frac{g'_{\pm}(\alpha_\pm)}{2(n+2)\alpha_\pm^{2n+1}}\right]d\tilde{y}\right]
         \exp\left[\int_{\tilde{y}_0^\pm}^y K_\pm^*(\tilde{y})d\tilde{y}\right]\\
         &=\left(\frac{y-\alpha_\pm}{{\tilde{y}_0^\pm}-\alpha_\pm}\right)^{\gamma_\pm+1}\exp\left[\int_{\tilde{y}_0^\pm}^y K^*_\pm(\tilde{y})d\tilde{y}\right],
        \end{align}
where the constant $\tilde{y}_0^\pm$ is chosen as $\tilde{y}_0^\pm\neq 0$ and $\tilde{y}_0^\pm<\alpha_\pm$, and 
	\be
         K^{*}_\pm(y)=K_\pm(y)+\frac{\gamma_\pm+1}{\left(y-\alpha_\pm\right)}.
        \ee
$K_\pm^*(y)$ is finite at $y=\alpha_\pm$ and the last integral in Eq.~(\ref{eq:redshift_generic}) has a finite value 
in the limit $y\rightarrow\alpha_\pm$.
The constant $k_0^{t,\pm}$ is related to $k_c^t := k^t(r=0)$ as
	\be\label{eq:k_ct}
         \frac{k_c^t}{k_0^{t,\pm}}=\exp\left[\int_{\tilde{y}_0^\pm}^\infty K_\pm(\tilde{y})d\tilde{y}\right].
        \ee
Combining Eqs.~(\ref{eq:redshift_generic}) and (\ref{eq:k_ct}), we find 
	\be
         k^t(y)=\left(\frac{y-\alpha_\pm}{{\tilde{y}_0^\pm}-\alpha_\pm}\right)^{\gamma_\pm+1}B(y)k_c^t,
        \ee
where 
	\be
         B(y)=\exp\left[\int_{\tilde{y}_0^\pm}^y \frac{\gamma_\pm+1}{\tilde{y}-\alpha_\pm}d\tilde{y}+\int_\infty^yK_\pm(\tilde{y})d\tilde{y}\right].
        \ee
        
Now, we shall consider a timelike observer who is rest at $r=\tilde{r}$ and will encounter the null ray twice, as done in~\cite{Miyamoto:2003wr}. The observed frequency is given by $\omega=-u_\mu k^\mu=k^t(y)$, where $u_\mu$ is the four-velocity of the observer.
We define $\omega_1=\lim_{y\rightarrow \alpha_-}k^t(y)$ and $\omega_2=\lim_{y\rightarrow \alpha_+}k^t(y)$.
Then, we obtain 
	\be\label{eq:redshift_LTB}
         \frac{\omega_2}{\omega_1}=\frac{\left({\tilde{y}_0^-}-\alpha_-\right)^{\gamma_-+1}}{\left({\tilde{y}_0^+}-\alpha_+\right)^{\gamma_++1}}
         \frac{B(\alpha_+)}{B(\alpha_-)}\frac{\left(y-\alpha_+\right)^{\gamma_++1}}{\left(y-\alpha_-\right)^{\gamma_-+1}}.
        \ee
From Eqs.~(\ref{eq:U1}), (\ref{eq:V1}) and (\ref{eq:g0}), we find
	\begin{align}
         y-\alpha_- =-\frac{U_0-U}{\Gamma_-},
        \;\;\;
         y-\alpha_+ =\alpha_+\left(-\frac{U_0-U}{\alpha_-\Gamma_-}\right)^\gamma
		\label{eq:y-alpha_p}.
        \end{align}
Substituting Eq.~(\ref{eq:y-alpha_p}) into Eq.~(\ref{eq:redshift_LTB}), 
the relation between $\omega_1$ and $\omega_2$ is written in terms of the outside coordinates $U$ and $V$,
	\be\label{eq:redshift_LTB2}
         \frac{\omega_2}{\omega_1}=\frac{\left({\tilde{y}_0^-}-\alpha_-\right)^{\gamma_-+1}}{\left({\tilde{y}_0^+}-\alpha_+\right)^{\gamma_++1}}
         \frac{B(\alpha_+)}{B(\alpha_-)}\left(\frac{\alpha_+}{\gamma\alpha_-}\right)^{\gamma_++1}\left(-\frac{U_0-U}{\Gamma_-}\right)^{\gamma-1}.
        \ee
Thus, because $\gamma-1>0$ from Eq.~(\ref{eq:agamma}), the redshift of the emitted particle diverges at the Cauchy horizon.

By the definition of the map of null rays $V=G(U)$, its derivative should be related to the redshift obtained above. By differentiating Eq.~(\ref{eq:g0}) with respect to $U$, we get 
	\be\label{eq:map_deriv}
         \frac{dV}{dU}=\frac{\gamma\alpha_+\Gamma_+}{\alpha_-\Gamma_-}\left(-\frac{U_0-U}{\alpha_-\Gamma_-}\right)^{\gamma-1}.
        \ee
Comparing Eq.~(\ref{eq:redshift_LTB2}) with Eq.~(\ref{eq:map_deriv}), we see that the power of the right-hand side in Eq.~(\ref{eq:redshift_LTB2}), $\gamma-1$, is identical to that in Eq.~(\ref{eq:map_deriv}).
From this fact, it seems that the divergence of the radiation power is related to the divergence of the redshift at the Cauchy horizon~\cite{Miyamoto:2003wr} (we will see below, however, that this is not the case in general).

\subsection{Degenerate case ($\zeta = \zeta_c$)}

In the degenerate case, $g_-(y)=0$ has the degenerate root at $y=\alpha_c$.
We subtract a pole of the integrand in Eq.~(\ref{eq:geodesic_equation}) (the expression of $k^t(y)/k^{t,-}_0$), 
which is second order in this case, 
	\begin{align}\label{eq:redshift_generic2}
         \nonumber \frac{k^t}{k_0^{t,-}}&=\exp\left[-\int^y_{\tilde{y}_0^-}\frac{4(n+2)\alpha_c^{2n+1}}{g_-''(\alpha_c)\left(\tilde{y}-\alpha_c\right)^2}d\tilde{y}\right]
         \exp\left[\int^y_{\tilde{y}_0^-} \widehat{K}_*(\tilde{y})d\tilde{y}\right]\\
         &=\exp\left[-\gamma_c\left(\frac{1}{y-\alpha_c}-\frac{1}{y_{\tilde{y}_0^-}-\alpha_c}\right)\right]\exp\left[\int^y_{\tilde{y}_0^-} \widehat{K}_*(\tilde{y})d\tilde{y}\right],
        \end{align}
where
	\be
         \widehat{K}_*(y)=K_\pm(y)+\frac{\gamma_c}{\left(y-\alpha_c\right)^2}.
        \ee	
$\widehat{K}_*(y)$ is finite at $y=\alpha_c$.
Combining Eqs.~(\ref{eq:redshift_generic2}) and (\ref{eq:k_ct}), we find 
	\be
         k^t(y)=\exp\left[-\frac{\gamma_c}{y-\alpha_c}\right]C(y)k_c^t,
        \ee
where 
	\be
         C(y)=\exp\left[-\frac{\gamma_c}{y_{\tilde{y}_0^-}-\alpha_c}+\int_{\tilde{y}_0^-}^y \frac{\gamma_c}{\left(\tilde{y}-\alpha_c\right)^2}d\tilde{y}+\int_\infty^yK_-(\tilde{y})d\tilde{y}\right].
        \ee
        
Let us consider a timelike observer who is rest at $r=\tilde{r}$ and will encounter the null ray twice.
We define $\omega_{1,c}=\lim_{y\rightarrow \alpha_c}k^t(y)$ and $\omega_2=\lim_{y\rightarrow \alpha_+}k^t(y)$.
Then, we obtain 
	\be\label{eq:redshift_LTBD}
         \frac{\omega_2}{\omega_{1,c}}=\frac{\left(y-\alpha_+\right)^{\gamma_++1}}{\left({\tilde{y}_0^+}-\alpha_+\right)^{\gamma_++1}}
         \frac{B(\alpha_+)}{C(\alpha_c)}\exp\left[\frac{\gamma_c}{y-\alpha_c}\right].
        \ee
From Eqs.~(\ref{eq:V1}), (\ref{eq:U3}) and (\ref{eq:g2}), we find
	\begin{align}
         y-\alpha_c = -\frac{\tilde{U}_0-U}{\Gamma_c},
        \;\;\;
         y-\alpha_+ = \alpha_+\exp\left[\frac{\gamma_c}{\gamma_+\alpha_c}\right]\exp\left[\frac{\gamma_c\Gamma_c}{\gamma_+\left(\tilde{U}_0-U\right)}\right]\label{eq:y-alpha_p2}.
        \end{align}
Substituting Eq.~(\ref{eq:y-alpha_p2}) into Eq.~(\ref{eq:redshift_LTBD}), 
the relation between $\omega_{1,c}$ and $\omega_2$ is written in terms of the outside coordinates $U$ and $V$,
	\be\label{eq:redshift_LTBD2}
         \frac{\omega_2}{\omega_{1,c}}=\frac{\left(\alpha_+\right)^{\gamma_++1}\exp\left[\frac{\gamma_c(\gamma_++1)}{\gamma_+\alpha_c}\right]}{\left({\tilde{y}_0^+}-\alpha_+\right)^{\gamma_++1}}
         \frac{B(\alpha_+)}{C(\alpha_c)}\exp\left[\frac{\gamma_c\Gamma_c}{\gamma_+\left(\tilde{U}_0-U\right)}\right].
        \ee
Thus, because $\gamma_c\Gamma_c/\gamma_+<0$, the redshift of the emitted particle diverges at the Cauchy horizon.
To compare the redshift with the derivative of the map, we differentiate Eq.~(\ref{eq:g2}) with respect to $U$,
	\be\label{eq:map_deriv2}
         \frac{dV}{dU}=-\frac{\gamma_c\alpha_+\Gamma_c\Gamma_+}{\gamma_+A^{1/\gamma_+}}\frac{\exp\left[\frac{\gamma_c\Gamma_c}{\gamma_+\left(\tilde{U}_0-U\right)}\right]}{\left(\tilde{U}_0-U\right)^2}.
        \ee

Comparing Eqs.~(\ref{eq:redshift_LTBD2}) and (\ref{eq:map_deriv2}),  
we find that the redshift and $dV/dU$, which can be a definition of redshift, diverge in different ways. Unfortunately, we do not have any clear explanation for this discrepancy.



\section*{References}
\begin{thebibliography}{30}

  \bibitem{ArkaniHamed:1998rs}
  N.~Arkani-Hamed, S.~Dimopoulos and G.~R.~Dvali,
  Phys.\ Lett.\  B {\bf 429}, 263 (1998);
  I.~Antoniadis, N.~Arkani-Hamed, S.~Dimopoulos and G.~R.~Dvali,
  Phys.\ Lett.\  B {\bf 436}, 257 (1998);
  N.~Arkani-Hamed, S.~Dimopoulos and G.~R.~Dvali,
  Phys.\ Rev.\  D {\bf 59}, 086004 (1999).

\bibitem{Randall:1999ee}
  L.~Randall and R.~Sundrum,
  Phys.\ Rev.\ Lett.\  {\bf 83}, 3370 (1999);
  L.~Randall and R.~Sundrum,
  Phys.\ Rev.\ Lett.\  {\bf 83}, 4690 (1999).

\bibitem{Nakao_2010}
  K.~-i.~Nakao, T.~Harada, U.~Miyamoto,
  Phys.\ Rev.\ D {\bf 82}, 121501 (2010).
  
\bibitem{Okawa:2011fv} 
  H.~Okawa, K.~-i.~Nakao and M.~Shibata,
  Phys.\ Rev.\ D {\bf 83}, 121501 (2011).

\bibitem{Argyres:1998qn}
  P.~C.~Argyres, S.~Dimopoulos and J.~March-Russell,
  Phys.\ Lett.\  B {\bf 441}, 96 (1998);
  T.~Banks and W.~Fischler,
  arXiv:hep-th/9906038;
  R.~Emparan, G.~T.~Horowitz and R.~C.~Myers,
  Phys.\ Rev.\ Lett.\  {\bf 85}, 499 (2000);
  S.~B.~Giddings and S.~D.~Thomas,
  Phys.\ Rev.\  D {\bf 65}, 056010 (2002);
  S.~Dimopoulos and G.~L.~Landsberg,
  Phys.\ Rev.\ Lett.\  {\bf 87}, 161602 (2001).
 
 
  
 \bibitem{Hawking:1974sw}
  S.~W.~Hawking,
  Commun.\ Math.\ Phys.\  {\bf 43}, 199 (1975)
  [Erratum-ibid.\  {\bf 46}, 206 (1976)].

\bibitem{Kanti:2008eq}
  P.~Kanti,
  Lect.\ Notes Phys.\  {\bf 769}, 387 (2009).

\bibitem{Ford:1978ip}
  L.~H.~Ford and L.~Parker,
  Phys.\ Rev.\  D {\bf 17}, 1485 (1978).

\bibitem{Hiscock:1982pa}
  W.~A.~Hiscock, L.~G.~Williams and D.~M.~Eardley,
  Phys.\ Rev.\  D {\bf 26}, 751 (1982).

\bibitem{Brave_1998}
  S.~Barve, T.~P.~Singh, C.~Vaz and L.~Witten,
  Nucl.\ Phys.\  B {\bf 532}, 361 (1998).
  
\bibitem{Vaz:1998gd}
  C.~Vaz and L.~Witten,
  Phys.\ Lett.\  B {\bf 442}, 90 (1998);
  S.~Barve, T.~P.~Singh, C.~Vaz and L.~Witten,
  Phys.\ Rev.\  D {\bf 58}, 104018 (1998);
  T.~Harada, H.~Iguchi and K.~i.~Nakao,
  Phys.\ Rev.\  D {\bf 61}, 101502 (2000);
  T.~Harada, H.~Iguchi and K.~i.~Nakao,
  Phys.\ Rev.\  D {\bf 62}, 084037 (2000);
  T.~Tanaka and T.~P.~Singh,
  Phys.\ Rev.\  D {\bf 63}, 124021 (2001).
  
\bibitem{Singh:2000sp}
  T.~P.~Singh and C.~Vaz,
  Phys.\ Lett.\  B {\bf 481}, 74 (2000).

\bibitem{Harada:2000me}
  T.~Harada, H.~Iguchi, K.~i.~Nakao, T.~P.~Singh, T.~Tanaka and C.~Vaz,
  Phys.\ Rev.\  D {\bf 64}, 041501 (2001).

\bibitem{Miyamoto:2003wr}
  U.~Miyamoto and T.~Harada,
  Phys.\ Rev.\  D {\bf 69}, 104005 (2004).

\bibitem{Miyamoto:2004ba}
  U.~Miyamoto, H.~Maeda and T.~Harada,
  Prog.\ Theor.\ Phys.\  {\bf 113}, 513 (2005).

\bibitem{Harada:2001nj}
  T.~Harada, H.~Iguchi and K.~i.~Nakao,
  Prog.\ Theor.\ Phys.\  {\bf 107}, 449 (2002).
 
 \bibitem{Penrose:1969pc}
  R.~Penrose,
  Riv.\ Nuovo Cim.\  {\bf 1}, 252 (1969)
  [Gen.\ Rel.\ Grav.\  {\bf 34}, 1141 (2002)].

\bibitem{Miyamoto:2010vn}
  U.~Miyamoto, H.~Nemoto, M.~Shimano,
  Phys.\ Rev.\ D {\bf 83}, 084054 (2011).

\bibitem{Miyamoto_Nemoto_2011a} U. Miyamoto, H. Nemoto, and M. Shimano, Phys. Rev. D {\bf 84}, 064045 (2011).
 
 
 \bibitem{Ghosh_2001} S. G. Ghosh and A. Beesham, Phys. Rev. D {\bf 64}, 124005 (2001).
 \bibitem{B.Carr} B. J. Carr and A. A. Coley, Gen. Rel. Grav. {\bf 37} 2165 (2005).
 \bibitem{Myers_1986} R. C. Myers and M. J. Perry, Annals Phys. {\bf 172}, 304 (1986).
 
\bibitem{Joshi:1993zg} 
  P.~S.~Joshi and I.~H.~Dwivedi,
  Phys.\ Rev.\ D {\bf 47}, 5357 (1993).
\bibitem{BSW} M. Banados, J. Silk, and S. M. West, Phys. Rev. Lett. {\bf 103}, 111102 (2009).
\bibitem{Patil and Joshi et.al} M. Patil, P. J. Joshi, M. Kimura and K. Nakao, Phys. Rev. D {\bf 86}, 084023 (2012).
\bibitem{M. Bejger}M. Bejger, T. Piran, M. Abramowicz, and F. Hakanson, Phys. Rev. Lett. 109, 121101 (2012).
\bibitem{Z. k. Stuchlik}Z. k. Stuchlik and J. Schee, Class. Quantum Grav. 30, 075012 (2013).


 \bibitem{Giudice:2002}
 G.~F.~Giudice, R.~Rattazzi and J.~D.~Wells, Nucl. Phys. B {\bf 630}, 293 (2002).
 

\bibitem{Giddings:2004}
 S.~B.~Giddings and V.~S.~Rychkov, Phys. Rev. D {\bf 70}, 104026 (2004).
  
\bibitem{Hawking}
 S. W. Hawking and G. F. R. Ellis, \textit{Large Scale Structure of Spacetime} 
(Cambridge University Press, Cambridge, 1972).
 



  






  


  \end{thebibliography}
\end{document}